\pdfoutput=1
\documentclass[
showkeys,useAMS,twocolumn,nofootinbib,pra]{revtex4-1}
\usepackage{graphics, graphicx, amsmath, amssymb}
\usepackage{epstopdf}
\usepackage{dsfont}
\usepackage{hyperref}
\usepackage{color}
\usepackage[normalem]{ulem}

\newcommand{\eq}[1]{\begin{equation}\begin{aligned}#1\end{aligned}\end{equation}}
\newcommand{\iu}{\text{i}}
\newcommand{\eu}{\text{e}}
\newcommand{\ha}{\hat{a}}
\newcommand{\had}{\hat{a}^\dagger}

\newcommand{\ket}[1]{\left|#1\right\rangle}
\newcommand{\bra}[1]{\left\langle#1\right|}
\newcommand{\braket}[2]{\left.\left\langle#1\right|#2\right\rangle}

\newcommand{\vac}{\ket{\text{vac}}}

\begin{document}
\title{Entanglement generation via diffraction}
\date{\today}
\author{Aaron Z. Goldberg}
\email{goldberg@physics.utoronto.ca}
\affiliation{Department of Physics, University of Toronto, Toronto, ON, M5S 1A7}
\author{Daniel F. V. James}
\affiliation{Department of Physics, University of Toronto, Toronto, ON, M5S 1A7}
\begin{abstract}
Quantum entanglement is an important resource for next-generation technologies. We show that diffracting systems can supplant beam splitters, and more generally interferometric networks, for entanglement generation --- systems as simple as screens with pinholes can create entanglement. We then discuss the necessary and sufficient conditions for entanglement to be generated by states input to any passive linear interferometric network. Entanglement generated in free space can now be harnessed in quantum-optical applications ranging from quantum computation and communication to quantum metrology.
\end{abstract}
\maketitle

That interacting quantum systems can become entangled \cite{Schrodinger1935} enables numerous applications \cite{NielsenChuang2000,Horodeckietal2009}. Entangled photons, for example, can be used to implement quantum computers \cite{Milburn1989,Knilletal2001,Edamatsu2007,OBrien2007}. Advantages in teleportation \cite{Bennettetal1993,Bouwmeesteretal1997,Boschietal1998}, cryptography \cite{Gisinetal2002,Curtyetal2004,AolitaWalborn2007,Tangetal2014,Baumletal2018}, and metrology \cite{Mitchelletal2004,PezzeSmerzi2009,Grossetal2010,Giovannettietal2011,GoldbergJames2018Euler} can also be unlocked by entangled photons. The ability to control the interactions between photons is thus a major requirement for modern technologies.

Entangled photons are usually generated using parametric down-conversion with a nonlinear crystal \cite{BurnhamWeinberg1970,Whiteetal1999}. This interaction, however, is nondeterministic, and the strengths of the desired nonlinearities limit experimental scalability \cite{OBrien2007}. An alternative method of inducing nonlinearities is by causing photons to interfere with each other at beam splitters \cite{FearnLoudon1987}, as shown in the famous Hong-Ou-Mandel experiment \cite{HongOuMandel1987}. Knill \textit{et al}. took advantage of this idea of using beam splitters to generate entanglement in their proposed scheme for quantum computation \cite{Knilletal2001}, paving the way for viable all-optical quantum computing \cite{OBrien2007}. Similarly, Fiur\'a\ifmmode \check{s}\else \v{s}\fi{}ek \cite{Fiurasek2002}, Zou \textit{et al}. \cite{Zouetal2001}, and Kok \textit{et al}. \cite{Koketal2002} showed how to use beam splitters to conditionally generate arbitrary entangled states of photons \cite{Fiurasek2002}, including those useful for cryptography \cite{Tangetal2014} and metrology \cite{Mitchelletal2004}. 

Here we suggest that the use of beam splitters vastly overcomplicates the infrastructure required for optical entanglement generation. Photons can be made to interfere in \textit{free space} following the principles of Fourier optics \cite{Goodman2005}, leading to entanglement by way of elementary diffraction. This can be used to enable entanglement-based technologies using rudimentary optical devices.

\begin{figure*}
	\includegraphics[width=\textwidth,trim= 0cm 1.6cm 0cm 1.6cm ,clip]{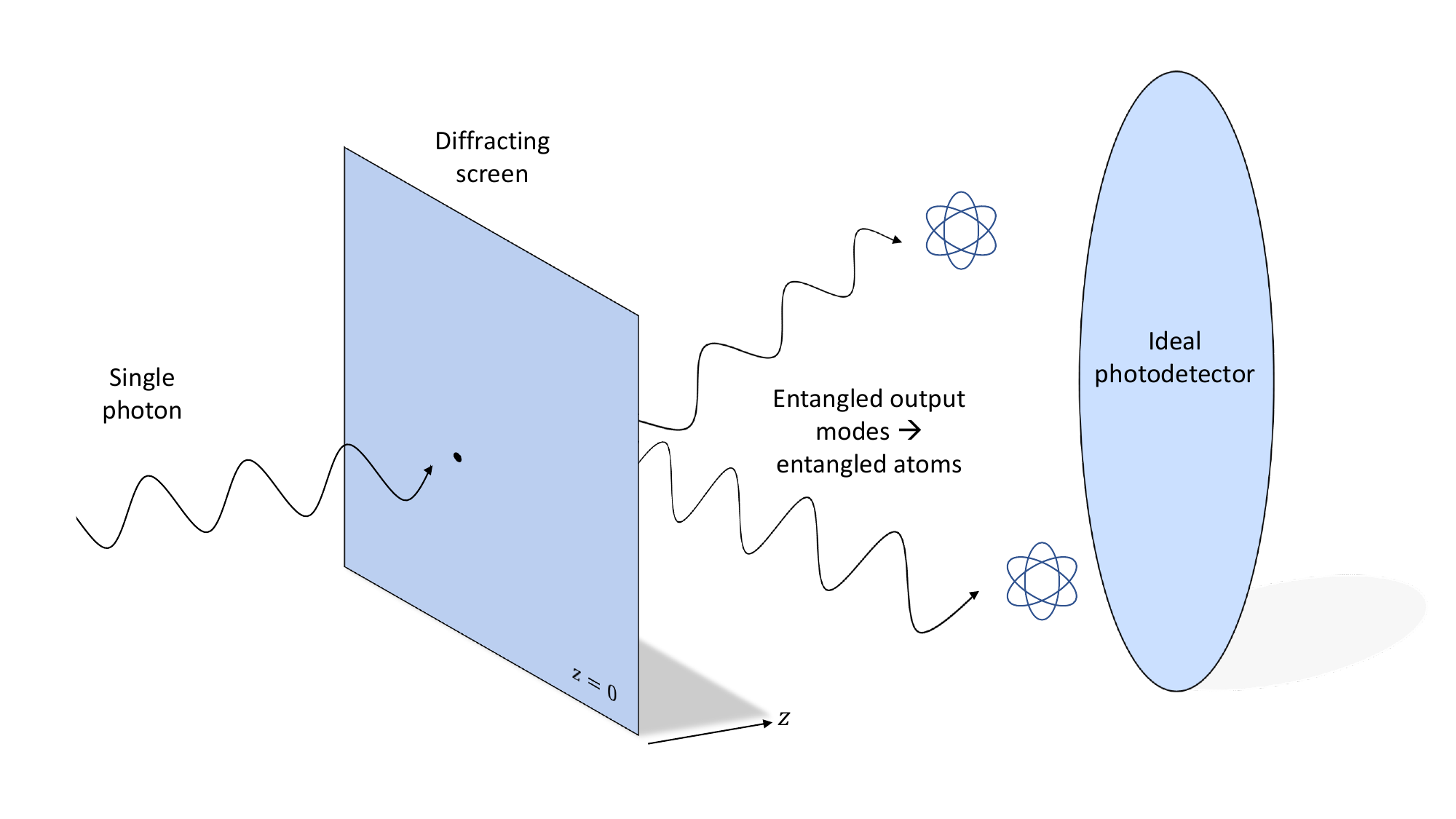}
	\caption{Schematic of entanglement generated via diffraction. A single photon mode impinges on a diffracting screen, such as an absorptive screen with a pinhole, resulting in output field modes that are entangled. Null detection of the photon leaves the atoms in an entangled state.
	}
	\label{fig:diffraction atoms}
\end{figure*}

Entanglement can seemingly be generated without interactions \cite{Krennetal2017}. We show that photons can become entangled using nothing more than diffraction --- without photon-photon interactions and without light-matter interactions. Consider a single photon diffracting through a pinhole, with two atoms placed between the pinhole and a perfect detector (Fig. \ref{fig:diffraction atoms}). Suppose further that the photon energy can resonantly drive a transition between the atomic ground and excited states. Remarkably, if the photon does not arrive at the detector, a scenario reminiscent of interaction-free measurement  \cite{ElitzurVaidman1993}, the two atoms are projected into an entangled state. We investigate the entangling properties of diffraction for generating useful quantum states of light for near-term applications.

As an electric field $E\left(\mathbf{r}\right)$ propagates through a lens, an aperture, or free space, it transforms via an impulse response function $h$ that depends on the geometry of the setup, through \cite{Goodman2005}
\eq{
	E_\text{out}\left(\mathbf{r}\right)=\int d\mathbf{r_0}\, h\left(\mathbf{r},\mathbf{r_0}\right)E_\text{in}\left(\mathbf{r_0}\right).
	\label{eq:electric field transformation impulse}
} 
This is mathematically equivalent to a beam splitter: 
a set of orthonormal modes $\left\{\mathcal{E}_m\left(\mathbf{r}\right)\right\}$ transforms to a superposition of output modes as $\mathcal{E}_m\left(\mathbf{r}\right)\to\sum_n U_{mn}\mathcal{E}_n\left(\mathbf{r}\right)$, for unitary matrix \eq{U_{mn}=\int d\mathbf{r}\,d\mathbf{r_0}\,\mathcal{E}^*_n\left(\mathbf{r}\right)h\left(\mathbf{r},\mathbf{r_0}\right)\mathcal{E}_m\left(\mathbf{r}\right).\label{eq:unitary matrix from h}}
The mode transformations \eqref{eq:electric field transformation impulse}-\eqref{eq:unitary matrix from h} have been explicitly calculated for numerous systems; for example, a Gaussian beam diffracting through a circular aperture transforms into a linear combination of Laguerre-Gaussian beams \cite{SchellTyras1971,BellandCrenn1982,Tanakaetal1985}. 

However, this has all been done with classical light. A quantized version of this transformation allows us to speak of input-output relations for the operators $\had_m$ that create excitations in mode $\mathcal{E}_m$, via the relation
\eq{\had_m\to\sum_n U_{mn}\had_n\label{eq:input-output creation operators}.} In other physical contexts, this type of \textit{mathematical} transformation has been shown to create entanglement between the output modes for the vast majority of nonclassical input states \cite{Jiangetal2013,GoldbergJames2018nonclassical}; here we discuss how it applies to this \textit{physical} transformation. A single photon $\ket{1}_0$ in a Gaussian beam, for example, can transform into the entangled state $U_{00}\ket{1}_0\ket{0}_1+U_{01}\ket{0}_0\ket{1}_1$, where $\ket{n}_i$ describes a state of $n$ photons in a Laguerre-Gaussian mode labelled by $i$. In this second-quantized sense of modal occupation numbers, the output photons will similarly be entangled given most nonclassical states of input photons.

The choice of basis in which to investigate modal entanglement depends on computational simplicity and experimental discernibility.
It is straightforward to see the entanglement generated in the plane wave basis.
A single photon with wavenumber $k$ and occupying the plane wave mode travelling in the $\mathbf{n}$ direction is represented by $\ket{1}_\mathbf{n}\equiv \had_{\mathbf{n}}\vac$, and is associated with the electric field $E\left(\mathbf{r}\right)\propto \eu^{\iu k\mathbf{n}\cdot\mathbf{r}}$, where $\mathbf{r}=\left(x,y,z\right)$. Other, more general electric fields can be written as 
\eq{E\left(\mathbf{r}\right)\propto \int d\Omega_{\mathbf{n}}\, \phi\left(\mathbf{n}\right)\eu^{\iu k\mathbf{n}\cdot\mathbf{r}}\label{eq:electric field mode},}
associated with annihilation operators $\had_\phi=\int d\Omega_{\mathbf{n}}\, \phi\left(\mathbf{n}\right) \had_{\mathbf{n}}$, where one component of $\mathbf{n}$ is allowed to be imaginary to account for evanescent waves \cite{Setalaetal1999,BornWolf1999}.
The mode functions $\phi_\text{in}\left(\mathbf{n}\right)$ and $\phi_\text{out}\left(\mathbf{n}\right)$ give the amplitudes for finding the incoming and outgoing electric fields in plane- or evanescent-wave mode $\mathbf{n}$, and are found by taking Fourier transforms of the incoming and outgoing electric fields. They respectively specify which modes are entangled with each other at the input and output.

As a simple example of entanglement generation via diffraction, consider a paraxial beam travelling in the $+z$-direction, impinging on a diffracting screen at $z=0$ (Fig. \ref{fig:diffraction atoms}). The transformation at this screen is well-described by a response function $h\left(\mathbf{r},\mathbf{r_0}\right)=\delta(\mathbf{r}-\mathbf{r_0})M\left(\mathbf{r}\right)$, providing a mask function $M\left(x,y\right)$ that achieves  \eq{E_\text{out}\left(x,y,0\right)=M\left(x,y\right)E_\text{in}\left(x,y,0\right).} 
Defining the Fourier transform of this mask function by $\large\tilde{M}\left(f_x,f_y\right)=\int dx\,dy\,M\left(x,y\right)\eu^{-\iu\left(xf_x+yf_y\right)}$, we find that \eq{
	\frac{\phi_\text{out}\left(\mathbf{n^\prime}\right)}{\left|kn_z^\prime\right|}=\int d\Omega_{\mathbf{n}}\,\phi_\text{in}\left(\mathbf{n}\right)\tilde{M}\left[k\left(n_x^\prime-n_x\right),k\left(n_y^\prime-n_y\right)\right].
	\label{eq:f_out any mask}
} 
This means that various choices of mask functions will entangle specific output modes for a given input mode $\phi_\text{in}$. A delta-function $\phi_\text{in}$ implies that the input is a plane wave, and the output plane wave modes are entangled for any $\phi_\text{out}$ that is not a delta function.

One mask function that exemplifies entanglement generation is the cosine grating 
\eq{M\left(x,y\right)=\tfrac{\sqrt{2}}{\left|k\left(u_z+n_z\right)\right|}\cos\left(u_xx+u_yy\right)}
 for unit vector $\mathbf{u}$, which has been used in optical intensity reconstruction \cite{FribergFischer1994}. This mask transforms an input plane wave $\phi_\text{in}(\mathbf{\tilde{n}})=\delta\left(\mathbf{\tilde{n}}-\mathbf{n}\right)$ to a superposition of plane waves $\phi_\text{out}\left(\mathbf{\tilde{n}}\right)=\left[\delta\left(\mathbf{\tilde{n}}-\mathbf{n}-\mathbf{u}\right)+\delta\left(\mathbf{\tilde{n}}+\mathbf{n}+\mathbf{u}\right)\right]/\sqrt{2}$, which enacts a special case of \eqref{eq:input-output creation operators}
\eq{
	\had_{\mathbf{n}}\to \frac{\had_{\mathbf{m}}+\had_{\mathbf{-m}}}{\sqrt{2}}, \quad \mathbf{m}\equiv \mathbf{n}+\mathbf{u}.
}
This transformation achieves entanglement between plane waves modes $\mathbf{m}$ and $-\mathbf{m}$:
\eq{\ket{1}_\mathbf{n}&\to\tfrac{1}{\sqrt{2}}\ket{1}_\mathbf{m}\otimes \ket{0}_\mathbf{-m}+\tfrac{1}{\sqrt{2}}\ket{0}_\mathbf{m}\otimes \ket{1}_\mathbf{-m}\\
	\ket{N}_\mathbf{n}&\to\sum_{j=0}^{N}\sqrt{\binom{N}{j}}\ket{j}_\mathbf{m}\otimes\ket{N-j}_\mathbf{-m}
	,\quad\text{etc.}
\label{eq:plane wave entanglement}}
We see that a single photon input in a plane wave mode leads to modal entanglement between output plane waves (see Ref. \cite{vanEnk2005} for a discussion of single-particle entanglement), as do most nonclassical input states, including all Fock states $\ket{N}_\mathbf{n}$ \cite{Jiangetal2013}. Similar, arbitrary transformations can be obtained by appropriately tailoring the mask functions $M$. Diffraction can thus be readily implemented instead of a beam splitter for entanglement generation.

The entangled photons created with \eqref{eq:plane wave entanglement} can be used in the thought experiment from Fig. \ref{fig:diffraction atoms}. If a single photon $\ket{1}_\mathbf{n}$ is incident on such a screen and a pair of ground-state atoms with a transition resonant with the photon energy are placed in the path of output modes $\mathbf{m}$ and $-\mathbf{m}$, unsuccessful detection of the photon projects the pair of atoms onto a Bell state. This can immediately be used for quantum computation and quantum communication protocols in which Bell states are the crucial resource. The idea of using undetected photons for imaging has been previously considered \cite{Lahirietal2015}, showing the power of such spatially-entangled photons. Other entangled states with different coefficients can be created by appropriately altering the mask function $M$.

These results can easily be extended to multimode cases, in which diffraction is similarly able to entangle multiple output modes for separable inputs. Consider the same plane wave impinging on a circular aperture of radius $R$. The mask function is given by $M\left(\mathbf{x}\right)\propto\Theta\left(R^2-\mathbf{x}\cdot\mathbf{x}\right)$, where $\Theta$ is the Heaviside step function, yielding
\eq{
	\phi_\text{out}(\mathbf{n^\prime})
	\propto\left|\frac{n_z^\prime}{n_z}\right|\text{jinc}\left(R\left|k\right|\sqrt{\left(n_x-n_x^\prime\right)^2+\left(n_y-n_y^\prime\right)^2}\right),	
	\label{eq:bessel fout}
}
where $\text{jinc}(x)=J_1(x)/x$ and $J_1(x)$ is the Bessel function of the first kind. A continuum of output plane wave modes $\mathbf{n^\prime}$ are coupled to by the initial plane wave, with amplitudes given by \eqref{eq:bessel fout}:
\eq{\had_{\mathbf{n}}\to\int d\Omega_{\mathbf{n^\prime}} \phi_\text{out}\left(\mathbf{n^\prime}\right)\had_{\mathbf{n^\prime}}.\label{eq:continous input output}} Each photon incident in mode $\mathbf{n}$ is output in a superposition over modes $\mathbf{n}^\prime$ with amplitudes given by $\phi_\text{out}\left(\mathbf{n}^\prime\right)$. Just like in the Hong-Ou-Mandel effect, nonclassical states of photons incident in this single mode will cause the output photons to be modally entangled. Diffraction can thus be used to replace beam splitters. 

A modal decomposition that is more amenable to experimental techniques involves Gaussian beams. As mentioned previously, a Gaussian beam with nonclassical photon statistics incident on a circular aperture yields an outgoing beam with entanglement between Laguerre-Gaussian modes, which has also been studied by Ref. \cite{Xiaoetal2017}. This entanglement can be harnessed using the vast array of classical techniques for controlling structured light \cite{Forbes2017}, including recent results for mode sorting without postselection \cite{Mirhosseinietal2013,Zhouetal2017,Fontaineetal2019}. The notion of quantum-mechanical transformations at apertures has been probed in Ref. \cite{Buseetal2018}, where it was shown that quantum entanglement can be \textit{preserved} between photons passing through a sub-wavelength circular aperture\textcolor{black}{, without any significant hindrance due to photon loss or other dissipative effects that might stem from the system being open \cite{PlenioKnight1998}. These results are especially promising due to the general lack of noise in photonic systems \cite{OBrien2007}}. Diffraction by way of metamaterials \cite{Wangetal2018}  has been proven capable of generating entanglement, and some transformations using diffraction at double slits \cite{Sadanaetal2019} have been investigated. These show the viability of using arbitrary mode transformations for achieving \textcolor{black}{experimentally-}useful quantum entanglement.

One may establish an arbitrary transformation $E_\text{in}\to E_\text{out}$ by tailoring an appropriate impulse response function. Defining the two-dimensional Fourier transform in the $z=0$ plane by $\mathcal{F}$, the convolution theorem applied to  \eqref{eq:electric field transformation impulse} yields
\eq{
h\left(\mathbf{r}-\mathbf{r_0}\right)=\mathcal{F}^{-1}\left[\frac{\mathcal{F}\left(E_\text{out}\right)}{\mathcal{F}\left(E_\text{in}\right)}\right].
\label{eq:convolutions solution}
}
This can immediately be used to entangle a desired set of orthogonal modes. For example,  \eqref{eq:convolutions solution} can be readily calculated for any desired transformation $\mathcal{E}_m\to \sum_n U_{mn}\mathcal{E}_n$ when the mode functions are Hermite-Gaussian modes; when $\mathcal{E}_m$ is the fundamental Gaussian mode, the function $h\left(\mathbf{r}-\mathbf{r_0}\right)$ is a Gaussian function multiplied by a polynomial. A suitably-constructed function $h\left(\mathbf{r}-\mathbf{r_0}\right)$ will thus entangle the output modes $\left\{\mathcal{E}_n\right\}$ for the correct nonclassical input states in mode $\mathcal{E}_m$.

We next discuss the requirements for input states to yield entangled output states, which \textcolor{black}{have} not been discussed in the context of diffraction. 
Reference \cite{Jiangetal2013} \textcolor{black}{studied} the condition\textcolor{black}{s} for a finite-dimensional multimode transformation on pure states to leave the output state fully separable between \textit{all} of the modes\textcolor{black}{:\eq{\ket{\psi_1}_1\otimes\ket{\psi_2}_2\otimes\cdots\ket{\psi_N}_N\to\ket{\phi_1}_1\otimes\ket{\phi_2}_2\otimes\cdots\ket{\phi_N}_N.} 
	One can express this requirement in the language of creation operators for each mode by using the Fock-Bargmann representation to identify an analytic function $B\left(z_1,\cdots,z_N\right)$ on $\mathds{C}^N$ with every $N$-mode pure state $\ket{\Psi}$
	\cite{Bargmann1961,Segal1962,Bargmann1962,Glauber1963}:
	\eq{&\ket{\Psi}=B\left(\had_1,\cdots,\had_N\right)\vac.}
	In this language, states that generate separable outputs correspond to products of functions of creation operators for each mode that transform under \eqref{eq:input-output creation operators} to products of new functions of creation operators for each mode: \eq{B\left(\mathbf{z}\right)=B_1\left(z_1\right)\cdots B_N\left(z_n\right)\to \tilde{B}_1\left(z_1\right)\cdots \tilde{B}_N\left(z_N\right),} without any mixing between the modes such as $\tilde{B}\left(z_i,z_j\right)\neq \tilde{B}_i\left(z_i\right)\tilde{B}_j\left(z_j\right)$. Such a transformation } happens if and only if the input state is a classical state, i.e., a multimode coherent state \textcolor{black}{with $B_i\left(z\right)\propto \exp\left(\alpha_i z\right)$ for all $i$}\cite{Glauber1963,Sperling2016}, or a particular type of squeezed state with equal squeezing in each mode \textcolor{black}{[$B_i\left(z\right)\propto \exp\left(-\lambda z\vphantom{a}^2\right)$ for all $i$]}. All other input states yield entangled outputs. 

\textcolor{black}{While the mathematical results of \cite{Jiangetal2013} explain how to generate entanglement between \textit{some} output modes, they don't specify which modes will be entangled. We thus extend these results to give necessary and sufficient conditions for generating entanglement between \textit{specific} output modes. For example, we can ask what input states yield output states that remain separable between the first two modes, such as
\eq{\ket{\psi_1}_1\otimes\ket{\psi_2}_2\otimes\cdots\ket{\psi_N}_N\to\ket{\phi_1}_1\otimes\ket{\phi_2}_2\otimes\ket{\Phi}_{3,\cdots,N}.\\
} The conditions are again related to coherent and squeezed states, with some important generalizations.
} 

\textcolor{black}{We answer this question in Appendix \ref{sec:app two mode entanglement} and summarize the main results here. The key insight of \cite{Jiangetal2013} was recognizing that separable modes have separable functions $B$ that are products of functions for each mode, and thus inspecting whether the logarithm $G\left(\mathbf{z}\right)=\ln B\left(\mathbf{z}\right)$ separates into sums of functions for each mode to is equivalent to identifying whether those modes are separable. Further, unitary transformations given by \eqref{eq:input-output creation operators} effect the transformation $\mathbf{z}\to \mathbf{U}\mathbf{z}$, where the unitary matrix $\mathbf{U}$ has elements $U_{mn}$. Comparing the series expansions of $G\left(\mathbf{z}\right)$ and $G\left(\mathbf{U}\mathbf{z}\right)$ conclusively tells us whether a given state will generate entanglement under a given transformation.} 

 \textcolor{black}{A subset of output modes will be separable if and only if the input modes that couple to them are associated with functions (Appendix \ref{sec:app two mode entanglement}) \eq{B_i\left(z\right)\propto\exp\left(\alpha_i z-\lambda z^2\right).} Other input modes $m$ that couple to none of the desired output modes $n$, i.e., modes for which $U_{mn}=0$ for all of the specified output modes $n$, have no restrictions on their functions $B_m\left(z\right)$. } 

\textcolor{black}{In the language of quantum states, }when inspecting a specific subset of output modes, coupled to by a set of input modes $\left\{\ha_1,\cdots,\ha_M \right\}$, the output modes are separable if and only if the input state is of the form
\eq{
	\ket{\Psi}=\hat{D}\left(\boldsymbol{\alpha}\right)\hat{S}_1\left(\lambda\right)\cdots \hat{S}_M\left(\lambda\right){B}\left(\had_{M+1},\cdots\right)\vac.
	\label{eq:states that generate no two mode entanglement}
}
Here, $\hat{D}\left(\boldsymbol{\alpha}\right)=\prod_{i=1}^{M}\eu^{\alpha_i\had_i-\alpha_i^*\ha_i}$ is the $M$-mode displacement operator that creates multimode coherent states when acting on the vacuum, $\hat{S}_i\left(\lambda\right)=\eu^{\lambda\left(\had_i\vphantom{a}^2-\ha_i^2\right)/2}$ is the squeeze operator for the $i^\text{th}$ mode (we take the squeezing strength $\lambda$ to be real without loss of generality by appropriately rephasing the input and output modes), and ${B}\left(\had_{M+1},\cdots\right)=B_1\left(\had_{M+1}\right)B_2\left(\had_{M+2}\right)\cdots$ is any separable function of operators for the remaining input modes. Equation \eqref{eq:states that generate no two mode entanglement} is the necessary and sufficient condition for an input state to generate no entanglement at a multimode interferometric network, of which diffraction is our current example.

By inspecting  \eqref{eq:states that generate no two mode entanglement} we see that classicality of the input modes $1$ to $M$ is the important consideration for whether entanglement will be generated between a specific subset of output modes. \textcolor{black}{Classical states $\hat{D}\left(\boldsymbol{\alpha}\right)\vac$ generate no entanglement, and \textit{most} other states generate entanglement.} The only nonclassical states that generate separable outputs in the particular $n$ output modes need \textit{every} input mode that couples to these output modes to be squeezed by the same amount\textcolor{black}{, such as $\hat{S}_1\left(\lambda\right)\cdots \hat{S}_M\left(\lambda\right)\vac$}; all other separable input states \textcolor{black}{that are nonclassical} will generate entangled outputs. 

\textcolor{black}{The conditions for generating separable outputs limit the sets of input modes that can be unoccupied.}  An input state that is squeezed in a few modes but unpopulated in other modes that \textit{could} couple to the specific output modes will still generate entanglement in the output modes\textcolor{black}{, because it has $B_i(z)\propto\exp(-\lambda z^2)$ for the first few modes but $B_j(z)=0$ for the others. For nonclassical states to generate separable outputs, even some of the input vacuum modes must be squeezed, thus the conditions for a nonclassical state to generate no entanglement are rather stringent. In contrast, if some of the vacuum modes that couple to the specified output modes are empty, only classical states will generate separable outputs.} Since the vacuum state of each mode is a coherent state, coherent state inputs suffice to generate separable outputs, as expected from their classicality.

In the continuum case [e.g.,  \eqref{eq:bessel fout}-\eqref{eq:continous input output}], an infinite number of modes may couple to every subset of output modes. \textcolor{black}{This is seen by the continuum of nonzero coefficients in the mode functions $\phi\left(\mathbf{n}\right)$.} All of these modes would need to be squeezed by the same amount to achieve the particular type of squeezed state that generates no entanglement at beam splitters [per \eqref{eq:states that generate no two mode entanglement}], which is infeasible in practice. \textcolor{black}{A nonclassical state in the continuum case would thus need an infinite number of equally-squeezed modes to generate separable outputs.} 

In such a situation, the only \textit{feasible} input states that generate no entanglement are coherent states, because these can have the other input modes in their vacuum states\textcolor{black}{. All practically-feasible nonclassical input states, squeezed in a finite number of modes, }  will generate entanglement. Along the lines of the assertion by Refs. \cite{Holleczeketal2011,Gabrieletal2011} that all nonclassical states yield quantum entanglement, we thus posit that the only \textit{physically realizable} states that generate no entanglement in the continuum case are coherent states \textcolor{black}{\eq{\ket{\Psi}=\hat{D}\left(\boldsymbol{\alpha}\right)\vac.}} All other input states will yield entangled outputs, even from setups as simple as diffraction.

\textcolor{black}{The discussion so far has focused solely on pure quantum states. In the context of mixed states, there exist another set of two-mode nonclassical states that generate no entanglement when undergoing the mode transformations \eqref{eq:input-output creation operators}. This set comprises the SU(2)-unpolarized states, formed from arbitrary incoherent sums of projections onto different photon-number subspaces, which can be both nonclassical and non-Gaussian \cite{GoldbergJames2018nonclassical}. In the case of $N>2$ modes, only some such sums of projectors are separable, of which the set of thermally-occupied modes at identical temperatures 
\eq{\rho(T)=\sum_{m,n,\cdots p=0}^{\infty} \eu^{-\frac{m+n+\cdots p}{k_BT}}\ket{m}\bra{m}\otimes\ket{n}\bra{n}\otimes\cdots\ket{p}\bra{p} }	
	is the only known example. As in \cite{GoldbergJames2018nonclassical}, arbitrary multimode displacements combined with equal multimode squeezing of this state will also produce states that generate no entanglement. The continuum case again requires that these properties hold for an infinite number of modes. It may be possible for this continuous set of modes to be thermally-occupied at similar temperatures (the state will generate entanglement if the temperatures are not \textit{identical}). However, a combination of thermal states is classical, so the only known nonclassical states in N>2 modes that generate no entanglement are those requiring the squeezing conditions from above, which we have already discussed as being prohibitive. This strongly suggests that all experimentally-relevant nonclassical states will generate entanglement via diffraction in both the pure- and mixed-state scenarios.}

The photons output from this type of protocol display true quantum entanglement. The spatial modes occupied at the output are distinguishable, allowing for nonlocal interactions between the modes. Although our initial examples focused on plane wave modes, our treatment can be extended to any set of orthogonal modes using  \eqref{eq:convolutions solution}, including those carrying orbital angular momentum (OAM) such as Laguerre-Gaussian modes. OAM can further be converted to spin angular momentum using $q$-plates \cite{Marruccietal2006,Marruccietal2011} and other straightforward systems \cite{Bliokhetal2011}, creating nonlocal polarization entanglement between photons.

A final note regarding the types of entanglement that can be created by passive linear optics is warranted. Generalized beam splitters can enact arbitrary unitary transformations
\eqref{eq:input-output creation operators} on the input creation operators, which is not equivalent to arbitrary unitary operators acting on the quantum states; these photon-number-conserving operators are restricted to $SU(N)$ transformations. The entangled states that can be created with passive linear optics alone are thus a subset of all entangled states (as shown explicitly in Ref. \cite{MoyanoFernandezGarciaEscartin2017}). In the case of a two-mode transformation
, for example, the output states from an arbitrary $N$-photon separable input state $\ket{\psi_\text{in}}=\ket{m}_a\otimes\ket{N-m}_b$ are parametrized by the four parameters $m$, $n\equiv N-m$, $\theta$, and $\phi$:
\eq{\ket{\psi_\text{out}}\propto &\sum_{k=0}^m\sum_{l=0}^n\binom{m}{k}\binom{n}{l}\sqrt{\left(n+k-l\right)!\left(m+l-k\right)!}\\
&\times\cos^{k+l}\frac{\theta}{2}\sin^{m+n-k-l}\frac{\theta}{2}\eu^{\iu\left(m-n+l-k\right)}\left(-1\right)^{n-l}\\
&\times\ket{n+k-l}_a\otimes\ket{m+l-k}_b;} the ``NOON'' states $\ket{\psi_\text{out}}\propto \ket{N}_a\otimes\ket{0}_b+\ket{0}_a\otimes\ket{N}_b$ are only included in this set for $N=2$ total photons. 		The generation of multiphoton GHZ-type states, as another example, cannot be achieved in this way. 
Extra processing steps could be used to convert the entangled states generated by diffraction to other desired entangled states. 

\textit{Conclusions}: 
We have shown that entanglement can be generated using the impulse response functions of simple optical devices. Since optical devices act like beam splitters by unitarily transforming electromagnetic field modes, they can be used in other applications requiring interference of various modes of light, ranging from interaction-free measurement to boson sampling. Diffracting screens are but examples of systems that can be tailored for achieving arbitrary beam-splitter-like transformations; similar results are achievable using the entirety of Fourier optics. This remarkably straightforward technique for controlling effective photon-photon interactions can be immediately implemented in the vast array of quantum optical experiments that utilize beam splitters.

Due to the challenge of characterizing mixed state entanglement, our discussion of the necessary and sufficient conditions for entanglement generation was \textcolor{black}{mainly} restricted to pure states. Some progress has been made regarding the conditions for bipartite mixed states to generate entanglement \cite{GoldbergJames2018nonclassical}; we hope to extend this to the multimode case in the near future. We again speculate that all \textit{physically realizable} nonclassical states impinging on a diffracting screen will lead to entangled output modes, even when the inputs are mixed states.

\begin{acknowledgments}
	This work was supported by the NSERC Discovery Award Fund \#480483 and by the Alexander Graham Bell Scholarship \#504825.
\end{acknowledgments}

\appendix
	\section{}
\label{sec:app two mode entanglement}

Here we show the necessary and sufficient conditions for a finite-dimensional unitary transformation of a separable $N$-mode input state to yield an $N$-mode output state in which a particular subset of output modes is separable. The condition is that all of the input modes that couple to the particular output subset must be squeezed coherent states, with equal magnitudes of squeezing in these input modes. This extends the results of Ref. \cite{Jiangetal2013}, who showed that equal squeezing in every input mode is a necessary and sufficient condition for all output modes to be fully separable.

Following Ref. \cite{Jiangetal2013}, we consider \textit{connected} unitary transformations $\ha_i\to\sum_{k=1}^{N}\ha_kU_{kj}$ for orthogonal modes $\ha_j$ and unitary matrix $U$. Connectivity dictates that the unitary transformation cannot be decomposed into a set of disconnected transformations on the subset modes, implying that $\left|U_{kj}\right|<1\, \forall k,j$. For our general transformations, our unitary matrices are sizeable; one must consider all of the output modes coupled to by the specific input modes, all of the other input modes that can lead to those output modes, the other output modes coupled to by the extra input modes, and so on. Having equal squeezing in all $N$ of these input modes is a large restriction on the separable inputs that will generate fully separable outputs; we investigate how this condition is relaxed when only a particular subset of output modes is to remain separable.

Using the Fock-Bargmann (or Segal-Bargmann) representation, we can represent every $N$-mode pure state by an analytic function on $\mathds{C}^N$
\cite{Bargmann1961,Segal1962,Bargmann1962,Glauber1963}:
\eq{&\ket{\Psi}=B\left(\had_1,\cdots,\had_N\right)\vac,\\
	B\left(z_1,\cdots,z_N\right)&=\sum_{n_1,\cdots,n_N}\frac{\braket{n_1,\cdots,n_N}{\Psi}}{\sqrt{n_1!\cdots n_N!}}z_1^{n_1}\cdots z_N^{n_N}.}
The unitary transformation effects $B_\text{out}\left(\mathbf{z}\right)=B_\text{in}\left(U\mathbf{z}\right)$ for $\mathbf{z}\equiv\left(z_1,\cdots,z_N\right)$. To gain insight into the separability of the input and output states, we consider the function $G\left(\mathbf{z}\right)=\ln\left[B\left(\mathbf{z}\right)\right]$, which for a pure product input state takes the form \eq{G_\text{in}\left(\mathbf{z}\right)=\sum_{k=1}^{N}G_k\left(z_k\right).} For a particular $n$ output modes to remain separable, we require that \eq{G_\text{out}\left(\mathbf{z}\right)=\sum_{k=1}^{n}\tilde{G}_k\left(z_k\right) + \tilde{G}\left({z_{n+1},\cdots,z_N}\right).} We will analyze the conditions for this separability to hold, given the relation $G_\text{out}\left(\mathbf{z}\right)=G_\text{in}\left(U\mathbf{z}\right)$.

We start by taking the Maclaurin expansion of the input and output functions:
\eq{G_\text{in}&=\sum_{d=0}^\infty \sum_{j=1}^{N}\lambda_j^{(d)}z_j^d} for input functions $G_j(z)=\sum_{d=0}^{\infty}\lambda_j^{(d)}z^d$, and similarly \eq{
	G_\text{out}(\mathbf{z})&=\left(\sum_{d=0}^\infty\sum_{j=1}^n \xi_j^{(d)}z_j^d\right)+\tilde{G}\left(z_{n+1},\cdots,z_N\right)\\
	&=\sum_{d=0}^\infty \sum_{j=1}^{N}\lambda_j^{(d)}\left(\sum_{k=1}^{N}U_{jk}z_k\right)^d.
}
This Maclaurin expansion can be made valid, even when $B_\text{in}\left(\mathbf{0}\right)=0$, by appropriately displacing the vacuum and leaving entanglement properties unchanged \cite{Jiangetal2013}.

The above equations must match order by order in $d$. We are looking for conditions on the input state expansion coefficients $\left\{
\lambda_k^{(d)}\right\}$ that will generate output states with no entanglement between modes 1 through $n$.

The $d=0$ conditions can always be satisfied for any $0^\text{th}$-order input coefficients $\lambda_k^{(0)}$:
\eq{
	\sum_{k=1}^N\lambda_k^{(0)}=\sum_{k=1}^n\xi_k^{(0)}+\tilde{G}\left(0,\cdots,0\right).
}
The $d=1$ coefficients also always satisfy the separability requirements. They yield the relations
\eq{
	\sum_{j=1}^N \lambda_j^{(1)}U_{jk}=\xi_k^{(1)},\quad k\in\left\{1,\cdots,n\right\}
}
which simply specify the particular output coherent states in modes 1 through $n$ that would be generated by input coherent states.

To create output states that are separable between modes 1 through $n$, the second-order conditions require that no terms of the form $z_kz_{k^\prime},\,k^\prime\neq k$ be created, where again $k\in\left\{1,\cdots,n\right\}$. This yields the conditions
\eq{
	\sum_{j=1}^N\lambda_j^{(2)} U_{jk}U_{jk^\prime}=\xi_k^{(2)}\delta_{k,k^\prime}\quad\forall\, k\in\left\{1,\cdots,n\right\}\quad\forall\,k^\prime.
}
Then, rewriting the right hand side using the unitarity condition $\delta_{k,k^\prime}=\sum_{j=1}^N U_{jk}^*U_{jk^\prime}$, we can rearrange to find the conditions
\eq{
	\sum_{j=1}^N c_{jk} U_{jk^\prime}&=0\quad\forall\, k\in\left\{1,\cdots,n\right\}\quad\forall\,k^\prime\\
	c_{jk}&\equiv\left(\lambda_j^{(2)} U_{jk}-\xi_k^{(2)}U_{jk}^*\right).
}
This implies that vectors formed by $\mathbf{c}_k=(c_{1k},\cdots,c_{Nk})$ are orthogonal to all $N$ rows of the unitary matrix $U$, and so the former must be zero vectors; i.e.,
\eq{
	\lambda_j^{(2)}U_{jk}=\xi_k^{(2)}U_{jk}^*, \quad k\in\left\{1,\cdots,n\right\},
}
which immediately yields our desired result. For input modes labelled by $j$ that do not connect to output modes 1 through $n$ ($U_{jk}=0$ for $k\in\left\{1,\cdots,n\right\}$), any squeezing $\lambda_j^{(2)}$ can still generate separable output modes 1 through $n$. For input modes that do couple to output modes 1 through $n$,
\eq{
	\left|\lambda_j^{(2)}\right|=\left|\xi_k^{(2)}\right|,\quad \forall\, j\quad\forall\,  k\in\left\{1,\cdots,n\right\},
}
implying that all of these input modes must be squeezed by the same amount.

\subsection{Satisfying the $d>2$ conditions.}
The order $d>2$ equations are much more difficult to satisfy. For generating fully separable output states, one can use matrix norms to show that the $d>2$ equations can never be satisfied, thus requiring $\lambda_k^{(d)}=0$ for $d>2$ \cite{Jiangetal2013}. We next show that the relaxed condition of $n$-mode separability at the output requires that only the specific input modes $\ha_j$ that couple to the subset of $n$ output modes must have $\lambda_j^{(d)}=0$ for all $d>2$. 

The conditions for separability of output modes 1 through $n$ are ($j_1\in\left\{1,\cdots,n\right\}$)
\eq{
	\xi_{j_1}^{(d)}\delta_{j_1,j_2}\cdots\delta_{j_1,j_d}=\sum_{k=1}^N\lambda_k^{(d)} U_{kj_1}\cdots U_{kj_d}.
}
The same orthogonality condition as before yields \eq{
	\xi_{j_1}^{(d)}U_{k j_1}^*\delta_{j_1,j_3}\cdots \delta_{j_1,j_d}=\lambda_k^{(d)}U_{kj_1}U_{kj_3}\cdots U_{kj_d},
	\label{eq:higher order orthogonality condition}
}
where we have replaced $\delta_{j_1,j_2}$, rearranged, and set every coefficient of $U_{kj_2}$ to zero.

If there is no higher-order input term, i.e. $\lambda_k^{(d)}=0$, then  \eqref{eq:higher order orthogonality condition} is automatically solved. If input mode $k$ couples to none of output modes 1 through $n$, then $U_{kj_1}=0$ for $j_1\in\left\{1,\cdots,n\right\}$, and  \eqref{eq:higher order orthogonality condition} is again solved. However, when neither of the above conditions holds, i.e., when $\lambda_k^{(d)}\neq 0$ and $U_{kj_1}\neq 0$,  \eqref{eq:higher order orthogonality condition} cannot be solved. For example, when all of the indices $j_4,\cdots,j_d$ are equal to $j_3$, our equation can only be solved by $U_{kj_3}=0\quad\forall j_3\neq j_1$. Since there is more than one possible value for $j_1$ (entanglement is not a property of single-mode states), this condition implies $U_{kj_3}=0$ for all values of $j_3$. However, no row of $U$ can be identically zero, so our assumptions $\lambda_k^{(d)}\neq 0$ and $U_{kj_1}\neq 0$ cannot both hold. We thus find that the only way to satisfy  \eqref{eq:higher order orthogonality condition} is by forbidding $d>2$ terms for input modes that are connected to output modes 1 through $n$.

We have thus shown that the necessary and sufficient conditions for a fully separable $N$-mode input state to generate no entanglement between a particular $n$-mode subset of output modes is that all of the input modes coupling to those output modes are squeezed coherent states with equal squeezing. This includes all classical (coherent) states and some nonclassical (squeezed) states. All other nonclassical pure state inputs will generate entanglement in these output modes.

%


\end{document}